# Origin and Quantitative Control of Sertoli Cells

Lixin Feng[1], Yongguang Yang[2]

[1] Division of Human Genetics, Cincinnati Children's Hospital Medical Center, Cincinnati, Ohio, United States

[2] Department of Cancer Biology, University of Cincinnati, Cincinnati, Ohio, United States

Email: yangy9@mail.uc.edu

**Abstract**

Sertoli cell is the "nurse" in testes that regulates germ cell proliferation and differentiation. One Sertoli cell supports a certain number of germ cells during these processes. Thus, it is the determinant of male reproductive capability. Sertoli cells originate from the primitive gonads during embryonic stage and their proliferations continue throughout the pre-pubertal stage. The proliferation and final density of Sertoli cells in the testis are regulated by hormones and local factors through autocrine, paracrine as well as endocrine methods. In the concise minireview, the most recent progresses in the study of factors and signaling pathways that participate into regulating the proliferation and function of Sertoli cell were summarized.

**Keywords:**

## Introduction

Sertoli cells play a decisive role in the male sex determination during embryogenesis and spermatogenesis of adulthood [1]. A Sertoli cell can support the development of a group of germ cells that provide growth factors and physical support, and the number of Sertoli cells in the testes is closely related to testicular volume and sperm yield. Postnatal Sertoli cells undergo continuous proliferation and differentiation until puberty stops dividing and begins feeding the germ cells. Thus, the number of Sertoli cells directly determines the adult male fertility.

Endocrine hormones such as Follicle-stimulating hormone (FSH), estrogen, thyroxine and other paracrine growth factors including Insulin-like growth factors I (IGF-I) and Insulin-like growth factor II (IGF-II), Fibroblast growth factors (FGF), somatostatin C, activin, Transforming growth factor (TGF) and Interleukin (IL) 1, The proliferation is very important [2-6]. Glial-cell-derived neurotrophic factor (GDNF) proteins are members of the TGF-β superfamily protein, and their receptors GFRa1, RET and NCAM are expressed in testis, suggesting its involvement into testicular development and spermatogenesis [7-11]. Mouse testicular gene knockout, testicular specific overexpression as well as in vitro culture of spermatogonial stem cell (SSC) have demonstrated that a certain level of GDNF is necessary for the maintenance and proliferation of SSCs [12, 13]. Studies have further showed that GDNF and Follicle-stimulating hormone (FSH) can synergistically promote the proliferation of neonatal rat testicular Sertoli cells [14]; another study has also shown that GDNF itself can promote the embryonic development of mouse Sertoli cell proliferation [15].

## 1. Origin and development of Sertoli cells

Testes have two main functions: production of testosterone and formation of haploid germ cells (spermatogenesis)[1]. These functions are generally regulated by pituitary gonadotropin, which act on Leydig cells and seminiferous tubules that produce testosterone through luteinizing hormone (LH) and FSH, respectively through acting on interstitial tissues Sertoli cells.

Sertoli cell was first reported by Enrico Setoli in 1865, which plays an important role in the process of spermatogenesis [1]. By forming the constituent components of the seminiferous tubules, completely differentiated Sertoli cells form a tight junction between adjacent cells (blood testicular barrier) to control the flow of nutrients and growth factors to haploid germ cells [1]. Therefore, the so-called blood-testicular barrier is not the isolation between blood and testicular tissues, but precise boundary formed between haploid and diploid germ cells. Thus, a well-functioning Sertoli cell provides suitable mitogen, differentiation factor and energy for the germ cells during development, while protecting them from harmful components and removal of the host's own immune system [1, 16]. It is clear that the number and quality of spermatozoa are determined by the number of functional Sertoli cells. Thus, the size of the testes is usually used as a measure of the number of Sertoli cells [1].

The appearance of embryonic Sertoli cells in primitive gonads marks the beginning of embryonic testicular development [16-20]. These cells express the *Sry* gene, thus determining the male sex of the gonads [21-23]. In the mouse gonads, the *Sry* gene began to be expressed in somatic cells from day 10.5 (10.5 dpc) after mating (these somatic cells eventually differentiate into Sertoli cells), reach to a peak at 11.5 dpc and disappeared in the gonad of 12.5 dpc [24-26]. Thus, the expression of the *Sry* gene in Sertoli cells during sex differentiation is only last for less than 48 hours [27]. The *Sry* gene is first expressed in the middle of the gonad and then expands into the beak and tail, and its expression in each cell lasts about 8 hours [25, 28-31]; the human Sry gene is expressed at probably 41th days after ovulation and reach to its peak at 44th day [32]. Compared with mice, the expression of human Sry gene after sex determination is very low and also expressed in adult Sertoli cells [33]. The Sry gene of the pig began to express at 23th day after mating, and the expression is down-regulated from day 35$^{th}$ [34]. In sheep, the Sry gene can be detected after 23-44 days of mating [35]. The somatic cells expressing Sry gene are precursors of Sertoli cells and eventually differentiate into Sertoli cells. The progenitor Sertoli cells induce the development of testicular cells during the process of angiogenesis and development of testicular cord formation [36-40]; at the same time, it can induce the migration of muscle, endothelial and mesenchymal cells from the potential mesonephros to the gonadal [41, 42]. Sertoli cells provide signals for the formation of other cells, induce the differentiation of steroid precursor cells to form Leydig cells, promote germ cells into the mitotic resting period, leading to the removal of secondary renal tubular cells through apoptosis [18]. As the sole source of anti-Mullerian hormone, Sertoli cells are responsible for inhibiting the development of intrinsic female genitalia [1].

Immature Sertoli cells are distinctly different from mature Sertoli cells, both morphologically and biochemical functions. In addition to Sertoli cells, immature seminiferous tubules contain only pericytes and germline stem cells, that is also different from that of adults [1]. Since the differentiation of germ cell has not yet begun in this period, so the seminiferous tubules are solid without lumen. Approaching to adolescence, Sertoli cells gradually extend and establish a close connection between the cells. These cells began to produce semen; seminiferous tubules also become a cavity tube. The pattern of protein expression has also changed in differentiated Sertoli cells and began to produce factors such as transferrin [43] and inflammatory cytokines IL-1a [44-50]. Immature Sertoli cells continue to divide, but their proliferative capacity decreases with adolescence, and Sertoli cells cease to proliferation upon the formation of tight junction [51].

## 2. Function and quantity regulation of Sertoli cells in mouse testes

2.1. Sertoli cell function

The main function of Sertoli cells is to regulate spermatogenesis and to change the proportion of sperm production [52]. Specific functions of Sertoli cells include the provision of nutrients and support for germ cells, phagocytosis of degraded germ cells and debris, release of sperm and / or response to pituitary hormones to produce host proteins to regulate spermatogonial mitosis [16, 53-59].

A typical feature of Sertoli cells providing structural support for germ cells at the developmental stage is the formation of a tightly connected blood testes barrier between adjacent cells [52]. This barrier isolates the spermatogonia and the pre-leptotene spermatocytes in the basal part, which allows the pre-leptotene spermatocytes pass through to the lumen and form spermatids and sperm cells in the luminal cavity[60-62]. This structure assembles the formation of an immune barrier to isolate the highly differentiated germ cells (spermatocytes and sperms) from the host immune system, so that it does not stimulate the production of autoimmune reactions [52].

In the testes of adult male, the number of Sertoli cells determines the volume of the testes and the number of sperm produced per day [17]. Despite the differences among different species, the number of germ cells the supported by one Sertoli cell are fixed [17, 63]. Only immature Sertoli cells have proliferative capacity, so the number of Sertoli cells in each testis has been determined before adulthood [17]. There is significant difference among species in terms of the proliferation period of Sertoli cells. The proliferation of Sertoli cells in rodent mainly occur in embryonic and early stages after birth, however the proliferation macaque's Sertoli cells occur mainly before puberty stage [17]. However, as more data are obtained from more species, it is now possible to reach a consensus on the proliferation of Sertoli cells that the proliferation of Sertili cells of all species occurs mainly in embryos or neonatal and adolescent stages [17, 64].

Proper function of Sertoli cell is clearly necessary for normal differentiation of germ cells[1]. However, although many Sertoli cells expressed proteins have been identified, the effects of these factors on germ cell development are not clear in most cases[1]. The most convincing example of a single factor produced by Sertoli cells that plays a vital role in spermatogenesis is stem cell factor (SCF). SCF is specifically produced by Sertoli cells and their receptor c-Kit is expressed on spermatogonia [1]. Although the destruction of SCF or its receptor in mice leads to the absence of germ cells in the testes, the implication of this finding in human testes is unclear [1, 65]. Although Sertoli cells can be isolated and cultured through enzymatic digestion of the extracellular matrix *in vitro*, it is difficult to isolate Sertoli cells in the adult testes by enzymatic digestion with the close connection between the cells and the changes in cell membrane structure during adolescence [1]. *In vitro* studies on Sertoli cell function are focused on the effects of different endocrine and paracrine factors on the expression and activity of certain proteins such as transferrin, aromatase, androgen receptor (AR) and lactate Hydrogenase (LDHC), etc. [1].

In general, testes play their function under the control of host genes and gene products. Many regulatory factors function through acting on Sertoli cells. Here are some of the major hormones and regulatory factors.

## 2.2. Sertoli cell function and quantity regulation

### 2.2.1 Endocrine regulation

#### 2.1.1.1 Follicle-Stimulating Hormone (FSH)

FSH is known as the main endocrine hormone that regulates Sertoli cell function [1]. In testes, Sertoli cells are the only cells that express FSH receptors, and the combination of FSH and other factors is necessary for the proliferation of Sertoli cells [4, 66-68]. In addition, FSH stimulates aromatase activity and Sertoli cells to produce inhibin, lactic acid, transferrin and AR [69-72]. Individuals that lack functional FSH receptors have significantly smaller testis size than that in normal person; this is consistent with the observed decrease Sertoli cell numbers in FSH receptor deficient mice [73]. Despite the reproductive capacity of people lack of FSH receptors, the spermatogenesis is disordered (eg, oligozoopermia and teratozoopermia) [74].

#### 2.1.1.2 Thyroid hormone and thyroid stimulating hormone (Thyroid Hormone and Thyroid-Stimulating Hormone, TH and TSH)

A series of studies in experimental animals have shown that hypothyroidism leads to an extension of Sertoli cell proliferation, the differentiation of Sertoli cells was inhibited, Sertoli cell number and testicular volume were increased, and the number of spermatozoa was also increased [75, 76]. In contrast, high levels of T3 promote the differentiation of Sertoli cells, resulting in a smaller testis volume and a lower sperm count [1]. Although the effect of high levels of TSH has not been elucidated in details, previous study has shown that high levels of TSH interacted with FSH receptors [77]. Human hypothyroidism is associated with increased testicular volume (without puberty), which also indicates an increase in the number of Sertoli cells [78].

#### 2.1.1. 3 Prolactin (PRL)

A widely recognized fact is that, in both females and males, elevated levels of PRL inhibit gonadotropin secretion, thereby affecting reproductive capacity [1]. Sertoli cells express PRL receptors; PRL can stimulate the proliferation of Sertoli cells [79]. In rats and pigs, PRL promotes lactate secretion and protein synthesis [80]. In male mouse, specific destruction of PRL or its receptors has no effect on the reproductive capacity, suggesting that PRL does not play a major role in the maintenance of male reproductive health of male mice (for a review, see [81]). There is no report showing the relation of pathological changes of human testes with the mutations in PRL or its receptors.

### 2.1.1. 4 Growth Hormone (GH)

Convincing evidences have shown that systematically or locally produced GH and insulin-like growth factor (IGF) interferes with gonadal function [1]. Since Sertoli cells express GH receptors [82], systematic injection of GH into the boar increased the volume of Sertoli cells and promoted their maturation [83]. Failure to respond to GH response led to delayed sexual maturation but did not affect reproductive capacity [84]. In addition, the lack of GH might be associated with the development of smaller testes volumes; although this did not result in a decreased reproductive capacity, the number of Sertoli cells was reduced [85]. So far, no *in vitro* study has shown that these phenomena observed *in vivo* could reflect the physiological functions directly mediated by receptors on Sertoli cells. Although it has been reported that GH treatment with GH-free boys could lead to reduced testicular volume and functional damage to [86], other reports did not support this finding [87, 88].

### 2.3 paracrine regulation

### 2.3.1 Testosterone

Spermatogenesis relies on the presence of sufficient levels of intracellular testosterone [89]. AR is expressed by Sertoli cells but not germ cells [90], suggesting that androgen acts on the seminiferous tubules through the former. Recent gene knockout studies have shown that AR expression in Sertoli cells is necessary for the development of blood-testis barrier [91], meiosis, as well as normal development of germ cells after meiosis [92]. The expression level of AR in Sertoli cells is associated with its maturation[1]. Although the action of androgen on Sertoli cells can regulate its gene expression and proliferation [92-94], the importance of this effect on the early proliferation of Sertoli cells remains uncertain [92]. There have been reports of Sertoli cell tumors in patients with AR mutations *in vivo* [95], and Sertoli cells of these patients still exhibited immature properties even in the absence of tumors [96].

### 2.3.2 Estrogen

The role of estrogen in testes has been controversial in recent years. It was known that exposing male individuals before birth to estrogen would negatively affect their subsequent reproductive capacity [97]. However, the presence of estrogen receptors in testicular cells suggests that estrogen may has its own physiological function in the testes [98, 99]. Immature Sertoli cells produce estrogen through their own aromatase [100], whereas Leydig cells and germ cells become estrogen sources in adult testes [101]. Sertoli cells express estrogen receptor-beta

(ER-beta) throughout the maturation process. Thus, it may be possible to respond to estrogen very early stage [102]. Although there is no evidence showing the requirement of estrogen to exercise normal physiological functions for these cells, the function of male testes carrying nonfunctional estrogen receptors or aromatase is disrupted [103]. In addition, although male mice lacking functional aromatase activity have fertility at an early stage, they become infertile subsequently [104].

### 2.3.3 Growth Factors

A variety of growth factors have been isolated from the testes, including FGF-1, FGF-2, IGF-I, IGF-II, TGF-a, activin A and EGF[1]. The receptors for most of these growth factors are expressed on Sertoli cells [105-111]. Some of these factors, such as FGF-1, FGF-2, IGF-I, IGF-II, TGF-a, activin A and EGF stimulate the proliferation of Sertoli cells [67, 112-119]; while many others affect Sertoli cell functions such as the production of aromatase and transferrin (for a review, see [120]).

### 2.3.4 Inhibin B

Under the stimulation of FSH, Sertoli cells produce and secrete hormone peptide inhibin B into the cycling system[1]. It is generally believed that inhibin B inhibits pituitary production through endocrine pathway FSH and no clear evidence showing inhibin B acts on Sertoli cells through autocrine pathways[1]. The inhibin B in human blood is maintained at a high level at early stage after birth and slowly decline to a stable detectable level and gradually increases again at pubertal stage[1]. Clinically, the presence and function of Sertoli cells in childhood can be judged by analyzing the level of inhibin B [1]. In contrast, the level of adult inhibin B depends on the presence of germ cells, so it can reflect the entire spermatogenesis process [117, 121-123].

### 2.3.5 Glial cell-derived neurotrophic factor (GDNF)

As a member of the TGF-β superfamily, GDNF was first known for its ability to preserve and differentiate the primary dopaminergic neurons [124, 125]. It is also a potent growth factor for the culture of motor neurons *in vitro* [126-131]. Although the dopaminergic and spinal motor neurons were normal in GDNF $^{-/-}$ gene knockout mice, the development of kidneys were blocked and gastrointestinal couldn't be innervated and finally died in perinatal period [132-135]; the reproductive capacity GDNF $^{+/-}$ mice was reduced and eventually led to infertility [12]. In rat testes, the number of Sertoli cells was established two weeks before birth, and the expression of GDNF in the testis was the highest during this period [51, 136]. The results of nuclease protection showed that the expression level of GDNF gradually increased with the testicular development and reached to the peak at 1 week after birth, and decreased gradually in 2-3 weeks after birth, with the lowest level in adult testis [14]. The expression of GDNF was detected in Sertoli cells and TM4 cell lines at mRNA level, indicating that GDNF in the testes might be produced by Sertoli cells [9].

Although Sertoli cells were normal in both GDNF $^{+/-}$ and testis-specific expression mice, it has been reported that GDNF and FSH could act cooperatively to promot the proliferation of Sertoli cells in rat testicular lumps from 6day old rat [14]; another report also shows that GDNF can promote the proliferation of mouse Sertoli cells during embryonic development [15]. However, the molecular mechanism and signal transduction pathway of GDNF-promoting Sertoli cell proliferation are not yet clear and need further study.